\documentclass[aps,prb, twocolumn,amsmath,amssymb,superscriptaddress]{revtex4}

\usepackage{graphicx}
\usepackage{dcolumn,color}
\usepackage{bm}
\usepackage{hyperref}

\begin{document}

\title{Constraints on Jones transmission matrices from time-reversal invariance and discrete spatial symmetries}

\author{N.P. Armitage }
\affiliation{The Institute for Quantum Matter, Department of Physics and Astronomy,  Johns Hopkins University, Baltimore, MD 21218, USA}

\date{\today}

\begin{abstract}

Optical spectroscopies are most often used to probe dynamical correlations in materials, but they are also a probe of symmetry.   Polarization anisotropies are of course sensitive to structural anisotropies, but have been much less used as a probe of more exotic symmetry breakings in ordered states.  In this paper a Jones transfer matrix formalism is discussed to infer the existence of exotic broken symmetry states of matter from their electrodynamic response for a full complement of possible broken symmetries including reflection, rotation, rotation-reflection, inversion, and time-reversal.  A specific condition to distinguish the case of macroscopic time-reversal symmetry breaking is particularly important as in a dynamical experiment like optics, one must distinguish $reciprocity$ from time-reversal symmetry as dissipation violates strict time-reversal symmetry of an experiment.  Different forms of reciprocity can be distinguished, but only one is a sufficient (but not necessary) condition for macroscopic time-reversal symmetry breaking.  I show the constraints that a Jones matrix develops under the presence or absence of such symmetries.  These constraints typically appear in the form of an algebra relating matrix elements or overall constraints (transposition, unitarity, hermiticity, normality, etc.) on the form of the Jones matrix.  I work out a number of examples including the trivial case of a ferromagnet, and the less trivial cases of magnetoelectrics and vector and scalar spin ``chiral" states.    I show that the formalism can be used to demonstrate that Kerr rotation must be absent in time-reversal symmetric chiral materials.  The formalism here is discussed with an eye towards its use in time-domain THz spectroscopy in transmission, but with small modifications it is more generally applicable.
    
\end{abstract}

\maketitle

\section{Introduction}

Optical spectroscopies are used to study a wide range of phenomena in the field of correlated materials \cite{Basov11a}.  Generally, they have been used to probe the dynamical consequences of broken symmetry states of matter, fluctuations of an ordered state, and interactions between electrons.  Experiments like optical conductivity allow one to probe the relevant time and energy scales that characterize a state.  Typical examples of its use in the correlated electron field include it as a probe of the pseudogap state of high-temperature cuprate superconductors \cite{Basov05a}, mass renormalizations in heavy-fermion compounds \cite{Degiorgi99a,Bosse12a}, spectral weight redistribution in Mott insulators \cite{Molegraaf02a,Boris04a}, electron spin resonance of quantum magnets\cite{Katsumata09a,Morris14a}, and superconducting fluctuations \cite{Corson99a,Bilbro11a,Bilbro11b,Liu11a}.

Although techniques like Raman scattering \cite{Devereaux07a} and second-harmonic generation \cite{Kumar08a} have been applied to determining broken symmetries in ordered exotic states of matter, optical spectroscopies - where one is measuring the dipole excitations in linear response - have not been as commonly applied to directly determine broken symmetries.  Of course it is obvious that response functions of materials that depend on electronic structure are sensitive to structural anisotropies and so generally structural features manifest themselves as anisotropies in dynamical response.   Obvious examples are the $c$-axis transverse Josephson plasmon observed in the high temperature cuprate superconductors \cite{Tsvetkov98a} or the strongly anisotropic in-plane response in quasi-2D organic superconductors \cite{Kaiser10a}.  In these cases the states exhibit anisotropies that are mirrored in structural features.  However, it can also be the case that symmetries are broken in ordered systems, which are not directly tied to lattice anisotropies or structure.   In this case it can be a delicate matter in determining exactly what symmetries are broken in a particular ordered state.   A prominent example of this is the hidden ordered state in URu$_2$Si$_2$ compound whose resolution has withstood 20 years of measurements despite the fact it shows a dramatic signature in the heat capacity \cite{Bonn88a}.   The community should be motivated to develop new experimental tools to probe broken symmetries in correlated states of matter.  
 
The use of optical probes for determining new broken symmetries has recently become more prominent with the observation of a spontaneous Kerr rotation (rotation of reflected light) in a number of superconducting and correlated systems \cite{Kapitulnik09a}.   These experiments build on those from the early days of the cuprate  superconductor field where proposed parity (P) and time-reversal symmetry (TRS) violating ``anyon" phases for the pseudogap would have specific optical anisotropies  \cite{MarchRussell88a,Halperin89a,Levi91a,Dzyaloshinskii91a,Lawrence92a,Canright92a,Canright92b, Halperin92a,Shelankov92a}.  In recent experiments, the Kerr rotation in Sr$_2$RuO$_4$ that onsets at the superconducting T$_c$ has been interpreted as being consistent with a TRS breaking $p+ip'$ superconducting order parameter \cite{Xia06a}.   In the cuprates, a small but significant signal was seen at a temperature close to (but slightly below) that where the enigmatic pseudogap state of the cuprates onsets \cite{Xia08a,Karapetyan12a,Karapetyan13a}.   It was originally attributed to a ferromagnetic-type time-reversal symmetry (TRS) breaking  \cite{Xia08a}, and later proposed to be consistent with a chiral gyrotropic (TRS preserving) or magnetoelectric (TRS breaking) effects \cite{Hosur012a,Pershoguba13a,Orenstein11a,Aji12a}.   As discussed below and elsewhere, it is a matter of current debate whether or not a chiral material as such can exhibit a Kerr rotation \cite{Landau60a,Casimir66a,Halperin92a,Bungay93a,Bahar09a,Hosur14a,Cho14a}.  Another case where optics have been used to find new correlated states of matter is in the case of La$_{2-x}$Sr$_x$CuO$_4$, where a factor of two difference in the $a$ and $b$ direction optical conductivities was found in an underdoped crystal that had only a 1\% difference in the in-plane lattice constants \cite{Dumm03a}.   Although this material was orthorhombic to start with (so additional 1D correlations broke no additional symmetries) this electronic anisotropy was interpreted as being due to the occurrence of unidirectional ``stripe" correlations on top of orthorhombicity because the conductivity anisotropy onset was at approximately 80K despite the fact that the compound was orthorhombic at room temperature, the spectral weight along the more conductive direction was actually suppressed relative to the other (in contrast to expectations from the overlap integral), and the large scale of the effect.   Even more recently, \textcite{Lubashevsky14a} has found a small but significant polarization rotation that derives from an anomalous linear dichroism in thin films whose principle axes are not aligned along those of the crystal symmetry directions.  In  YBa$_2$Cu$_3$O$_y$ the effect has a temperature onset that mirrors the pseudogap temperature T$^*$ and is enhanced in magnitude in underdoped samples.    In $x=1/8$  La$_{2-x}$Ba$_{x}$CuO$_4$, the effect onsets above room temperature, but shows a dramatic enhancement near a temperature scale known to be associated with spin and charge ordered states.  These features are consistent with a loss of both C$_4$ rotation and mirror symmetry in the electronic structure of the CuO$_2$ planes in the pseudogap state.

In the use of optical spectroscopies to determine broken symmetry states of matter, it is important to have a well developed formalism for extracting the relevant information as it is easy for one symmetry to masquerade as another.   For instance, strong linear dichroism (absorption) can effectively rotate the plane of polarization of light, in a manner that could superficially be confused with a magnetic state with circular birefringence (phase retardation).   Both broken mirror and time-reversal symmetries can give circular birefringence.  It is important to precisely determine what aspects of the optical anisotropy correspond to TRS breaking as various kinds of chiral and gyrotropic orders can give similar experimental signatures.  The frequency scale of the optical experiment is also an important consideration.  Ordered states are generally defined in the $\omega \rightarrow 0$ limit.  This means that low frequency optical measurements may be very useful in determining the onset of a broken symmetry.   In this regard time-domain THz spectroscopy (TDTS) is a growing and important area of investigation, with important recent advances in THz polarimetry \cite{Castro05a,Morris012a,Neshat12a,Aschaffenburg12a,Castro12a,George12a,Nagashima13a}.  This technique generally satisfies the constraint in which these probes have frequency scales lower than the many frequency scales of interest.   It has the ability to measure complex response functions directly without needing to resort to Kramers-Kronig transform.   The fact that complex transmission functions are measured and radiation with well-defined polarization states used means that the simpler Jones matrix formalism can be used to analyze data.    A challenge with TDTS is performing it in reflectance where phase information is difficult to recover reliably.
 
 In this paper a formalism is discussed for inferring the existence of exotic broken symmetry states of matter from their electrodynamic response.  This formalism is discussed within the Jones transfer matrix approach and so is directly applicable to time-domain THz spectroscopy in transmission, but is more broadly applicable.  I discuss the consequences of discrete broken symmetries on ordered states of matter including the presence and absence of reflections, rotations, inversion, rotation-reflections and  time-reversal symmetry and the constraints they give on Jones matrices.   These constraints typically appear in the form of an algebra relating matrix elements or overall constraints (transposition, unitarity, hermiticity, normality, etc.) on the form of matrix.  This formalism is applied to other more exotic symmetry breakings such as various forms of time-reversal symmetry breaking.   As usual, the utility of symmetries is that one can still deduce quantitative and qualitative information even when the underlying equations of motion are unknown.

\section{General properties of Jones Matrices}

Due to its phase sensitivity, in TDTS one typically measures the complex transmission function and uses the Jones calculus to understand the interaction of light with a material system being investigated.  The complex amplitudes of the incident $E^i$ field  to the transmitted fields $E^t$ for propagation in a particular propagation direction can be related through a frequency-dependent $2\times2$ $``Jones"$ matrix.   In the most general case, the Jones matrix is comprised of 4 independent complex values.  In the basis of $x-y$ linear polarization it is

\begin{equation}
\hat{T} = \left[\begin{array}{cc}T_{xx} & T_{xy} \\T_{yx} & T_{yy}\end{array}\right] 
\left[\begin{array}{c}E_{x}^i \\E_{y}^i\end{array}\right]  = \left[\begin{array}{c}E_{x}^t \\E_{y}^t\end{array}\right].
\label{Tmatrix}
\end{equation}

The above equation forms an eigenvalue-eigenvector problem, where the Jones matrix has eigenvalues
 
 \begin{equation}
\kappa_{\pm} = \frac{1}{2} (T_{xx} + T_{yy} \pm \sqrt{ (T_{xx} - T_{yy} )^2  + 4 T_{xy}T_{yx} })
\label{eigenvalues}
\end{equation}

and (unnormalized) eigenvectors

\begin{equation}
\left[\begin{array}{c} 1 \\ \frac{  \kappa_{+} -  T_{xx} }{  T_{xy}}  \end{array}\right],  \medskip   \medskip   \medskip   \left[\begin{array}{c} 1 \\ \frac{  \kappa_{-} -  T_{xx} }{  T_{xy}}  \end{array}\right].
\label{eigenvectors}
\end{equation}
 
 If light is sent through the system under test with a well-defined linear polarization state in, for instance, the $x$ direction, then the Faraday rotation (rotation of transmitted light) may be measured through the relation tan$(\theta_F) = T_{xy} / T_{xx}$.   In principle, this is a complex quantity with real part of arctan$(\theta_F) $ equal to the Faraday rotation itself and the imaginary part of arctan$(\theta_F)$ equal to the ellipticity.   See Ref. \onlinecite{Morris012a} for further details.

For plane waves, the spatial and temporal dependence of the waves in free space is given in the usual way by multiplying these complex field amplitudes by a factor $e^{i(kz - \omega t)}$.  More complicated cases but still within the paraxial approximation can be treated by letting the complex amplitudes acquire a dependence on $x$ and $y$.  We envision a plane wave incident on a slab made of a particular material.   This slab has various symmetries that it inherits from both the particular point groups of the material as well the slab's macroscale structure.   For most of our analysis I assume the slab itself has the most generic shape that can be assumed without losing generality, which is a flat cylindrical solid with rotation $\hat{R}_z(\theta)$, $\hat{R}_y(\pi)$ and mirror $M_{xy}$, $M_{yz}$, and $M_{zx}$ symmetries.   Where specified I will also address the case where $M_{xy}$ symmetry is lost by additional features such as in the presence of a thin film on a substrate.  The presence of a substrate does not effect the consideration of symmetries that include only a single surface, such as $M_{xz}$ or $\hat{R}_z(\pi/2)$ , but can be important with symmetry considerations that involve ``backside" transmission (for example $M_{xy}$).  

Alternative to the Jones formalism the $4 \times 4$ matrix formalism of Berreman can be used \cite{Berreman72a}.  In this method one explicitly treats the time and spatially varying magnetic field as well.   Although there are disadvantages of the expanded dimensionality of the matrix, such a method can be useful in computing explicit optical properties of materials as one does not have to resurrect the two suppressed field variables with auxiliary equations when matching fields at boundaries.  It is particularly useful in calculating the properties of stratified media.  Still when only matters of symmetry are concerned the Jones matrix formalism is sufficient and I rely on it almost exclusively in the below discussion.

In TDTS a time limited pulse is sent through this slab and its amplitude and phase measured.   The electric field of a time-domain pulse $E_x(t)$ can be written as a superposition of Fourier components in the usual way.   For $x$ polarized light

\begin{equation}
\label{eq:fourier_comps}
E_x(t) = \frac{1}{\sqrt{2\pi}}\int_{-\infty}^{\infty} d \omega e^{i \omega t} E_x(\omega).
\end{equation}

Here, $E_x(t)$ is the purely real electric field propagating through the system, and $E_x(\omega)$ describes the amplitude of the complex Fourier components that comprise $E_x(t)$.   One can apply the Jones calculus to each of these Fourier components separately.  Going forward I will mostly replace the entries $T_{ij}$ in Eq. \ref{Tmatrix} with entries $A,B,C,D$ for convenience e.g. 

\begin{equation}
\left[\begin{array}{cc}T_{xx} & T_{xy} \\T_{yx} & T_{yy}\end{array}\right] =
\left[\begin{array}{cc}A & B \\C & D \end{array}\right] .
\label{ABCD}
\end{equation}

It is useful to be able to transform the $ \hat{ T}$ matrix into bases of different polarization states.   Let $\tilde{E^i}$ and $\tilde{E^t}$ be the electric field vectors in a transformed basis.   They are related to the electric fields in the original basis through the transformation matrix $E^i =  \hat{\Lambda }\tilde{E^i}$ and $E^t = \hat{ \Lambda} \tilde{E^t}$.   The transmission matrix in the transformed basis is then $\tilde{\hat{T}} = \hat{ \Lambda} ^{-1} \hat{T} \hat{ \Lambda}$.

One such matrix of practical importance is that of the transformation to a circular base in terms of left and right circularly polarized light
 
 \begin{equation}
\hat{\Lambda }= \frac{1}{\sqrt{2}}\left[\begin{array}{cc}1 & 1 \\i & -i \end{array}\right] , \; \hat{ \Lambda}^{-1} = \frac{1}{\sqrt{2}}\left[\begin{array}{cc}1 & -i \\1& i \end{array}\right].
\end{equation}
 
The transmission matrix in the circular basis in terms of the $x-y$ matrix elements is then
 
 \begin{equation}
\hat{T} _{circ}= \frac{1}{2} \left[\begin{array}{cc}A + D +  i (B-C)  & A - D - i (B+C) \\ A-D + i (B+ C) & A+D - i(B-C) \end{array}\right] 
\end{equation} 
 
\noindent which connects the transmission amplitudes of $R-$ and $L-$hand polarized light.  Other transformation matrices exist which represent other symmetry operations.   These will be discussed below.

\section{Constraints on Jones matrices from point symmetries}
\label{DiscreteSym}

Neumann's principle states that the symmetry transformations of any intrinsic physical property of the crystal must include $at$ $least$ the symmetry transformations of the point group of that crystal \cite{Nye85a}.   Therefore the symmetries that states of matter possess impose symmetry requirements on their Jones matrices and hence an algebraic relations among their elements.  Conversely, observed symmetries of Jones matrices and the algebraic relations between their elements (or the lack thereof) can tell us about symmetries that states of matter possess or break.   In this discussion, it is important to keep in mind that since Neumann's principle is usually only applied to intrinsic properties, its application to a macroscopic property like a Jones matrix formally requires that the system has a generic shape that itself has all relevant symmetries.   For the most general case, one can consider a flat cylindrical solid with rotation $\hat{R}_z( \theta)$, $\hat{R}_y(\pi)$ and mirror $M_{xy}$, $M_{yz}$, and $M_{zx}$ symmetries.  In practice, since in a TDTS measurement one measures a normalized transmission (e.g. transmission through a sample mounted to an aperture divided by transmission through a bare aperture), sample shape does not matter appreciably.  It may be appreciable only for samples where one lateral dimension is smaller or of order the wavelength and near field effects become relevant.

An advantage of the present approach is that it considers only symmetries and the directly measured experimental quantities e.g. ratios of transmitted electric fields to reference fields.   The difficulty with applying symmetry constraints to the intrinsic material constants (for instance the dielectric function) and then calculating the transmission is that this may involve a number of complications and model dependent assumptions.   If one were to calculate from the material constants, then for a particular symmetry one must write down the constrained form of the dielectric tensor $\hat{\epsilon}$, the magnetic permeability tensor $\hat{\mu}$, and the magnetoelectric susceptibility tensor $\hat{\alpha}$.   Then one must apply the boundary conditions and Fresnel equations properly in their full generality (e.g. accounting for magnetoelectric effects) to get the transmission matrix.  This is obviously much more involved than in the current approach.  Moreover for many complex materials the precise issue of electromagnetic boundary conditions can be complicated.   For instance, there is still ongoing discussion about the proper extension of the Fresnel equations to chiral materials \cite{Landau60a,Casimir66a,Halperin92a,Bungay93a,Bahar09a,Hosur14a,Cho14a}.   In the present approach, one avoids all this and concentrates on the experimentally measured quantities directly.   The precise form of $\hat{\epsilon}$, $\hat{\mu}$, or $\hat{\alpha}$ does not matter, nor does the nature of the boundary conditions.  The approach here borrows in part from a formalism for quasi-2D meta-material structures \cite{Menzel10a}.  Important differences to the present case of bulk broken symmetry states of matter will be discussed.

In general, the simplest examples come from rotation and reflection symmetries.   Rotation by an angle $\phi$ is represented by the transformation matrix

 \begin{equation}
\hat{R}_z(\phi) = \left[\begin{array}{cc} \mathrm{cos}{\phi} &  \mathrm{sin}{\phi}  \\ - \mathrm{sin}{\phi}  &  \mathrm{cos}{\phi}  \end{array}\right].
\end{equation}
 
 If a material possesses a $\hat{R}_z(\pi/ 2)$  rotation symmetry ($C_4$) around the $z$ axis, its Jones matrix will be invariant under a $\frac{\pi}{2}$ rotation.   Applying the relevant rotation matrix to Eq. \ref{ABCD} gives
 
  \begin{align}  
\hat{R}_z(\pi/ 2) &= \left[\begin{array}{cc} 0 &  1 \\  -1 &  0 \end{array}\right],  \nonumber \\    \tilde{\hat{T}} = \hat{R}_z^{-1} (\pi/2)  \hat{T}  \hat{R}_z(\pi/2) &= \left[\begin{array}{cc}D & -C \\-B & A \end{array}\right] .
\label{Piover2Rotation}
\end{align}

\noindent Hence the Jones matrix for a material with $\hat{R}_z(\pi/ 2)$ symmetry must have the form

\begin{equation}
C_4(z)  \;\;  \Longrightarrow \; \; \hat{T} = 
\left[\begin{array}{cc}A & B \\-B & A \end{array}\right].
\label{C4}
\end{equation}

If the material has a $\hat{R}_z(2\pi/3)$ rotation symmetry ($C_3$) as for instance encountered in rotations of cubic structures along their $[111]$ direction or in the corundum structure of materials like Cr$_2$O$_3$ around the $c$ axis, one has a Jones matrix of the form

\begin{equation}
C_3(z)  \;\;  \Longrightarrow \; \; \hat{T} = 
\left[\begin{array}{cc}A & B \\-B & A \end{array}\right].
\label{C3}
\end{equation}

$\hat{R}_z(\pi)$ rotation symmetry  ($C_2$)  imposes no constraints on the Jones matrix 
 
  \begin{align}
\hat{R}_z(\pi) &= \left[\begin{array}{cc} -1 &  0 \\  0 &  -1 \end{array}\right], \nonumber \\  C_2(z)  \;\;  \Longrightarrow \; \;  \tilde{\hat{T}}  = \hat{R}_z(\pi)^{-1}  \hat{T}  \hat{R}_z(\pi) &= \left[\begin{array}{cc}A & B \\C & D \end{array}\right] .
\label{PiRotation}
\end{align}

If an mirror symmetry exists with respect to the $xz$ plane

  \begin{align}
\hat{M}_{xz} &= \left[\begin{array}{cc} 1 &  0 \\  0 &  -1 \end{array}\right],  \nonumber \\  \tilde{\hat{T}}  = \hat{M}_{xz}^{-1}  \hat{T}  \hat{M}_{xz} &= \left[\begin{array}{cc}A & -B \\-C & D \end{array}\right],
\label{Mirror3}
\end{align}

\noindent and therefore the Jones matrix for a material with $\hat{M}_{xz}$ reflection symmetry (where $\hat{M}_{xz}$ is the matrix for reflections across the $xz$ plane) must be 

\begin{equation}
\hat{M}_{xz,yz} \;\;  \Longrightarrow \;\;  \hat{T} = 
\left[\begin{array}{cc}A & 0 \\ 0 & D \end{array}\right] .
\label{reflection}
\end{equation}

An identical form of the Jones matrix arises from $M_{yz}$ reflection symmetry as indicated above.  The  $\hat{T}$  matrix is diagonal if there exists any mirror plane parallel to the z axis and either the $\hat{x}$ or y axes.  One can see by the constraints in Eqs. \ref{C4} and \ref{reflection} that if a state has both $\hat{R}_z(\pi/ 2)$ rotation symmetry and a mirror symmetry along $x$ or $y$ axes then off-diagonal elements vanish and diagonal elements must be equal.

By simple rotation of Eq. \ref{reflection} one can show that if a mirror plane exists in any direction that contains the $z$ axis then the  $\hat{T}$  matrix will have the form

\begin{equation}
\hat{M}_{nz} \;\;  \Longrightarrow \;\;  \hat{T} = 
\left[\begin{array}{cc}A' & B' \\ B' & D' \end{array}\right] ,
\label{reflection1}
\end{equation}

\noindent where the elements are not independent, but are related by the trigonometric relations $A' =A \mathrm{cos}^2 \phi + D \mathrm{sin}^2 \phi$,  $D' =A \mathrm{sin}^2 \phi + D \mathrm{cos}^2 \phi$, and $B' =  (A - D ) \mathrm{sin}  \phi \: \mathrm{cos} \phi $ where $\phi$ is the rotation of the mirror plane with respect to the laboratory axes.    Eq. \ref{reflection} is a special case of \ref{reflection1} when a mirror plane is aligned along either $x$ or $y$ principle axes and the off-diagonal elements are zero.

Additional constraints can be made for materials that possess time-reversal symmetry such that their optical response can be said to be $reciprocal$.   Effects are non-reciprocal if they have opposite signs for the two states of a crystal that are related to each other by time-reversal.  Reciprocity and time-reversal symmetry is discussed in detail below.  In the remainder of this section I assume reciprocity which should be valid for all non-magnetic materials.   As will be shown below the $backwards$ transmission through a Jones matrix for a reciprocal material is equal to transpose of the Jones matrix through the forward direction e.g.

\begin{equation}
\hat{\underline{T}}^{b,-z} = \left[\begin{array}{cc}A & C \\B & D \end{array}\right] ,
\end{equation}

\noindent where the $\hat{\underline{T}}^{b,-z}$ refers to transmission through the ``backside" of the materials with wave propagation in the $-z$ direction and the underline here and below represents the presence of TRS.  Most useful for the typical experimental situation is the case where instead of considering backwards transmission through the sample, one considers forward transmission, but through a sample that has been rotated by $\pi$ around the $y$ axis.   In this case the Jones matrix is 

\begin{equation}
\hat{\underline{T}}^{b,+z} = \left[\begin{array}{cc}A & -C \\-B & D \end{array}\right] ,
\label{recip}
\end{equation}

\noindent where now  $\hat{\underline{T}}^{b,+z}$ refers to transmission through the backside of the materials with wave propagation in the $+z$ direction.  The minus signs in the off-diagonal part accounts for the fact that in the reference plane of the sample $x \rightarrow -x$ when performing the $\hat{R}_y(\pi)$  rotation.  Note that this expression is true $only$ for reciprocal systems.   We denote time-reversal invariance in this sense by the symbol $\underline{1}$.  For the most generic case, the forward going Jones matrix does not carry enough information to constrain the backwards propagation of the material completely.    I will discuss non-reciprocal and time-reversal symmetry broken systems in more detail below.  But for materials which are reciprocal, mirror symmetry in the $xy$ plane and Eq. \ref{recip} gives important constraints on the Jones matrices.   A mirror operation across the $xy$ plane is equivalent to a $\hat{R}_y(\pi)$ rotation followed by a mirror reflection across the $yz$ plane.  Therefore 

\begin{equation}
\tilde{\hat{\underline{T}}}  = \hat{M}_{yz}^{-1} \hat{R}_y(\pi)^{-1}  \hat{\underline{T}}  \hat{R}_y(\pi) \hat{M}_{yz} = \left[\begin{array}{cc}A & C \\B & D \end{array}\right] ,
 \label{Mxy}
\end{equation}

\noindent which gives a form for the   $\hat{T}$  matrix that is 

  \begin{equation}
\underline{1}, \hat{M}_{xy} \;\;  \Longrightarrow \;\;   \hat{T}   = \left[\begin{array}{cc}A & B \\B & D \end{array}\right] .
  \label{reflection2}
\end{equation}

Note that this expression is strictly invalid for thin films mounted to substrates as the substrate will violate the $ \hat{M}_{xy}$ symmetry.   In principle, the breaking of mirror symmetry by a substrate can be weak, but specific cases have to be analyzed individually.  Comparing Eqs. \ref{reflection1} and \ref{reflection2} one can see that the presence of a mirror symmetry along $any$ direction gives the same symmetric form for the Jones matrix (for a system with a TRS symmetric reciprocal response).

If a crystal has a 3D center of inversion ($P$ symmetry) and TRS then the Jones matrix also has the form of Eq. \ref{reflection2} as inversion is equivalent to a reflection across the $xy$ plane and $\hat{R}_z(\pi)$ rotation around $z$.   However, I have already shown in Eq. \ref{PiRotation} that symmetry under $\hat{R}_z(\pi)$ does not constrain the Jones matrix, therefore the constraint that reflection symmetry across $xy$ gives is the same as inversion.

Rotation-reflection symmetry with an improper axis $z$ can be handled through a combination of normal rotation (Eq. \ref{Piover2Rotation} for $ \hat{R}_z(\pi/ 2)$)  and the $\hat{M}_{xy}$ operation (Eq. \ref{Mxy}).   For $S_4$ and \underline{1} one gets the completely isotropic matrix

  \begin{equation}
\underline{1}, S_{4}(z) \;\;  \Longrightarrow \;\;   \hat{T}   = \left[\begin{array}{cc}A & 0 \\0 & A \end{array}\right] .
  \label{S4}
\end{equation}

Rotation-reflections with an  in-plane improper axis ($x$ or $y$) and \underline{1} give

  \begin{equation}
\underline{1},  S_{2}(x) \;\;  \Longrightarrow \;\;   \hat{T}   = \left[\begin{array}{cc}A & B \\B & D \end{array}\right] .
  \label{S4}
\end{equation}

Structures that cannot be mapped onto their mirror images by rotations and translations are chiral \cite{Barron04a,Simonet12a}.   Chirality itself generally does not put constraints on Jones matrices, but states that are chiral can be consistent with other symmetries that constrain the Jones matrices.   For instance, chiral reciprocal materials with TRS can have a $\hat{R}_x(\pi)$ symmetry

  \begin{equation}
\tilde{\hat{\underline{T}}} = \hat{R}_x^{-1} (\pi)  \hat{\underline{T}}  \hat{R}_x(\pi) = \left[\begin{array}{cc}A & -C \\-B & D \end{array}\right] ,
\end{equation}

\noindent such that 

\begin{equation}
\underline{1},  C_{2}(x) \;\;  \Longrightarrow \;\;  \hat{T} = 
 \left[\begin{array}{cc}A & B \\ -B & D \end{array}\right].
\label{C2x}
\end{equation}

Note that Eqs. \ref{reflection2} and \ref{C2x} are only compatible with each other for vanishing off-diagonal elements.   Therefore considering that circular optical activity is only possible when off-diagonal elements of the $\hat{T}$ matrix are anti-symmetric, one can see that to generate circular optical activity it is insufficient to break reflection symmetries only in the $xz$ and $yz$ planes, one must also break reflection symmetry in the $xy$ plane.   TRS chiral material can also exhibit a $\hat{R}_z(\pi/ 2)$ symmetry giving the constraint on the Jones matrix in Eq. \ref{C4}.   Chirality typically introduces anti-symmetric components to the dielectric tensor through an expansion first order in the light wavevector $k$.    This is consistent with the fact that in these TRS materials one can switch the sign of the off-diagonal components by considering transmission through the backside in which $ k \rightarrow -k$.   Also note that no reference to the ``screw axis" direction was made to derive Eq. \ref{C2x}, showing that even when transmission is perpendicular to the screw axis, chiral materials can show optical activity (although it will be of a complicated elliptical variety without  $C_{4}(z)$ symmetry).

Note that translational invariance itself imposes no constraints on Jones matrices.  In the context of our current treatment, this must be the case because plane waves are invariant in the direction perpendicular to their propagation direction.  But more generally it is only the point groups (crystallographic or magnetic) that determine the macroscopic symmetries properties of materials \cite{Dzyaloshinsky58a}.  Although at first glance trivial, this point is important when considering the properties of materials with space groups that have more complicated symmetries with glide or screw symmetries.  In these cases it is only the magnetic point group part of the global symmetry that imposes a constraint.

Note that domain structures can average out polarization anisotropies.  In order to see a broken symmetry optically, the symmetries must be broken globally over the entire beam spot.   Residual strain or stray field may provide an effective aligning field.  Alternatively, external fields need to be used to create monodomain samples.  These issues may be more pronounced in TDTS with its inherently long wavelengths and larger beam size.

\section{Time-reversal and Reciprocity}
\label{TimeReversal}

In any discussion on the constraints on Jones matrices from time-reversal symmetry it is important to define exactly what is meant by time-reversal symmetry as fortunately (or not) time marches inexorably in a single direction in all real experiments.   For plane waves traveling in the $+z$ direction the electric field can be written as $E_0 e^{i(kz - \omega t)}$ where $E_0$ is the two component vector in the $x$ and $y$ directions.   Time-reversal is trivially accomplished letting $t \rightarrow - t$ in this expression.   One can alternatively accomplish the equivalent of the time-reversal operation by the dual operation of complex conjugation of the amplitude $E \rightarrow E^*$ and letting $k \rightarrow -k$.  This can be seen simply by the fact that the physical electric field is $E_0 \; \mathrm{cos } (\omega t) = \frac{1}{2} (E_0 e^{i \omega t} + E_0^* e^{-i \omega t} )$.   Reversing the sign of the time ($t \rightarrow -t$) in this real physical electric field is mathematically equivalent to taking the complex conjugation of $E_0$.  That complex conjugation is necessary for the effective time-reversal operation can alternatively be see in the fact that it takes both it and the inversion of the wavevector to reverse the sign of Poynting's vector describing energy flow in an optical system.  Heuristically, conjugation can also be seen to be necessary to phase advance a time reversed wave sufficiently such the phase accumulated during time-reversed propagation is exactly canceled to reconstruct the initial wave with its initial phase relation.

Even time reversed waves from the dual operation of complex conjugation and wavevector inversion will not exactly reconstruct an original wavefront if a system has absorption or scattering (e.g. reflection).   Absorption and the generation of heat violates time-reversal invariance through the 2nd law of thermodynamics.   Since all real materials will show some dissipation (or reflection) at frequencies of interest, a strict form of time-reversal symmetry breaking of an optical process is generally not of interest to us.   This demonstrates that one must formulate a more applicable definition to be sensitive to time-reversal symmetry breaking in the state of matter itself.  This can be found in the concept of reciprocity \cite{Lorentz36a,Perrin42a,DeHoop59,Chandrasekhar60a,Shelankov92a,Potton04a}.   A system may be reciprocal if one can switch source and detector (with the possibility of additional constraints) and get the same result.   The importance of the reciprocity condition is in the fact that it is applicable even in the presence of absorption or scattering when other treatments lack ``true" time-reversal symmetry.  As in our discussion on point group symmetries, discrete time-reversal invariance (formulated in terms of reciprocity) puts certain important constraints on Jones matrices.  These were already exploited above in the discussion of how the Jones matrix behaves under  $\hat{R}_x(\pi)$ or  $\hat{M}_{xy}$ operations in non-magnetic materials.

A system can be said to be reciprocal if a response has the same sign for the two states of a material that are related to each other by time-reversal (e.g. time even).  A system is non-reciprocal if a response or component of the $\hat{T}$ matrix has the opposite sign for the time reversed state (time odd).   For further analysis we need a relation for the Jones matrix of a system in which the sense of time is reversed.   Although related formulations have been made previously \cite{DeHoop59,Shelankov92a,Potton04a}, a particularly simple treatment exists for the case of coherent polarized waves that can be treated with this Jones formalism.  Consider the situation when a wave $E^i$ is incident on a sample with a Jones matrix $ \hat{T} $.   The resulting wave must be analyzed by its projection on a reference wave $E^r$.  Channeling Onsager \cite{Onsager31a}, microscopic time-reversal invariance mandates that if both the sense of time in the experiment and the sample are reversed then the result must be the same e.g.

\begin{equation}
E^{r \dagger} \hat{T} E^i = E^{i *\dagger} (\overline{ \hat{T}}^{b,-z} ) E^{r*} , 
\end{equation}

\noindent where now the initial wave is the time-reversed $E^{r*}$, $ \overline{ \hat{T}}^{b,-z} $ is the Jones matrix for light propagation in the $-z$ direction through the backside of a time-reversed system, and the wave is analyzed by its projection on the time-reversed wave $E^{i *}$.  Here $\dagger $ signifies the Hermitian adjoint and the overbar represents the operation of time-reversal.  One can reverse the order of the inner product on the right side of this equation and take its complex conjugate to get

\begin{equation}
E^{r \dagger}  \hat{T} E^i =  E^{r \dagger}    (\overline{ \hat{T}}^{b,-z})^T  E^i.
\end{equation}

Relabeling the sense of time of the system on both sides of this equation one gets

\begin{equation}
\overline{ \hat{T}}   =   (  \hat{T}^{b,-z})^T.
\label{TRScondition}
\end{equation}

\noindent This gives the general result that the time-reversed Jones matrix equals the transpose of the Jones matrix for light propagating through the system's backside.  If the system has macroscopic TRS, then $\hat{T} =   \overline{ \hat{T}} $ and one arrives at the result used in Sec. \ref{DiscreteSym} that for a system that has a reciprocal response and hence macroscopic TRS $(\underline{1})$ the backside transmission is the transpose of the front side transmission

\begin{equation}
\underline{1} \;\;  \Longrightarrow \;\;    \underline{ \hat{T}}   =   ( \underline{  \hat{T}}^{b,-z})^T  .
\label{deHoopCondition}
\end{equation}

\noindent where again the underline reminds us that this is a relation for the Jones matrix of sample that has TRS.

A condition related to the above for interaction of an electromagnetic wave on a finite size scatterer based on a definition with incoming wave and a projection on a reference wave was first formulated by de Hoop \cite{DeHoop59,Potton04a} and is the most general expression of reciprocity.  The condition above is a representation in the Jones matrix formalism.   This $de$ $Hoop$ reciprocity distinguishes a form of time-reversal symmetry most applicable to real systems and experiments.  It is a sufficient (but not necessary) condition to say a state breaks TRS if it violates de Hoop reciprocity.  Not all TRS breaking states will show optical non-reciprocity.  It is necessary to break TRS macroscopically, which generally means that the combination of the time-reversal operation and a lattice translation is not a symmetry operation \cite{Dzyaloshinskii91a}.

Considering the above condition for de Hoop reciprocity, a measure of  time-reversal symmetry breaking is then the quantity $|T_{xy} - T_{yx}^{b,-z}|$ where  $T_{yx}^{b,-z}$ refers to transmission through the backside of the materials with wave propagation in the $-z$ direction.   As discussed above, more relevant to the usual experimental situation is the case where the waves are propagating in the $+z$ direction, but through a film that has been rotated around the $y$ axis by $\pi$.   In that case the measure of  symmetry breaking becomes $|T_{xy} + T_{yx}^{b,+z}|$ where the $ T_{yx}^{b,+z}$ refers to backside transmission but in a sample that has been flipped.  Examples of states that violate this condition of de Hoop reciprocity are discussed in Sec. \ref{Combined}.

In some systems it is possible to control the sense of the direction of time by cooling in a magnetic field (for a state like a ferromagnet) or dual magnetic and electric fields (for a magnetoelectric-like Cr$_2$O$_3$).  It is then a sufficient condition to say a state violates TRS if the equality

\begin{equation}
E^{r \dagger}  \hat{T} E^i = E^{r \dagger} \overline{ \hat{T} }E^i
\end{equation}

\noindent is not satisfied.  One can see that with this expression and Eq. \ref{TRScondition} that a necessary, but not sufficient condition for TRS will be an invariance of the off-diagonal components of $ \hat{T} $ with the time-reversal operation.

It is instructive to consider more restricted definitions of reciprocity.  For instance, consider the constraints implied by the hypothetical case where a time-reversed wave exactly reconstructs the initial wave.  Considering incoming and outgoing waves related by

\begin{equation}
 \hat{T}  E^i =  E^t.
  \label{Unitary1}
\end{equation}

If time-reversed version of the final wave can exactly reconstruct the initial wave and de Hoop reciprocity holds then 

\begin{equation}
 \hat{T}^T  E^{t*} =  E^{i*}
 \label{Unitary2}
\end{equation}

\noindent is true where the complex conjugation of $E^i$ and $E^t$ time reverses the wave amplitudes $\grave{a}$ $la$ the discussion above.   Taking the complex conjugate of Eq. \ref{Unitary2} and substituting Eq. \ref{Unitary1} into it for $E^t$  gives

\begin{equation}
[ \hat{T}^T]^*  \hat{T} E^i =  E^i
 \label{Unitary3}
\end{equation}

\noindent which can only be true if $\hat{T}$ is unitary.   This form of reciprocity is not particularly significant from a practical perspective because unitary matrices are norm preserving and hence preclude absorption or reflections.   Even purely dielectric media will induce reflections at interfaces, which destroys the unitary condition.   However, this condition does satisfy the simplest considerations for time-reversal symmetry and hence is important from a conceptual point of view.  One can call this form of time-reversal symmetry $unitary$ $reciprocity$ \cite{Potton04a}.

Another form of reciprocity can be determined by considering a situation where the initial wave $E^i$ is time-reversed, but made to propagate through the system from the ``backside."   The system could be said to be reciprocal in a restricted sense when the final waves in either case are time reversed versions of each other.     If the system is reciprocal in the sense of de Hoop and also reciprocal in this more restricted sense then a time reversed wave with polarization state $E^{i*}$ results in a backside transmitted wave $E^{t*}$ $via$

\begin{equation}
 \hat{T}^T  E^{i*} =  E^{t*}.
\end{equation}

\noindent Taking the complex conjugate of this expression gives

\begin{equation}
[ \hat{T}^{T}]^*  E^{i} =  E^{t}.
   \label{Hermitian2}
\end{equation}

Eqs. \ref{Unitary1} and \ref{Hermitian2} can only both be true if  $  \hat{T}  =   \hat{(T^{T})^*}   $ e.g.  $\hat{T} $ is Hermitian.   Hence one can call this kind of transmission properties $Hermitian$ $reciprocity$ \cite{Potton04a}.  Note that one should not confuse non-Hermiticity in the transmission matrix (or scattering matrices in general) with non-Hermiticity in the dielectric function. Non-Hermitian dielectric tensors are generally associated with bulk absorption, whereas non-Hermiticity in the $\hat{T} $ matrix (or scattering matrices more generally) are associated with phase delay.  The most obvious example of an optical component that violates Hermitian reciprocity is a quarter wave plate that turns linearly polarized light into circularly polarized light.  

Very asymmetric transmissions for forward and backward transmission will be obtained in the case where eigenpolarizations are non-orthogonal and the Jones matrix is non-diagonalizable \cite{Chipman94a,Tudor06b,Menzel10b}.   For this to be the case the Jones matrix needs to be non-normal e.g. $\hat{T}\hat{T}^\dagger \neq \hat{T}^\dagger \hat{T}$ where $\dagger$ signifies the Hermitian adjoint.   A diagonalizable $\hat{T} $ matrix for a sample that obeys de Hoop reciprocity will be similar for forward and backward going transmissions with identical transmission eigenvalues.  In contrast, one can see by inspection of Eqs. \ref{eigenvalues} and \ref{eigenvectors} that although the eigenvalues of a non-diagonalizable transposed backwards going Jones matrix are not different from the forward ones, the eigenvector associated with a particular eigenvalue will change its character.    It is a sufficient condition for the $\hat{T} $ matrix  to be normal that $T_{xy} = \pm T_{yx}$.  Symmetric transmission of the kind found in normal Jones matrices we can call $diagonal$ $reciprocity$.

A simple example of an optical component that breaks diagonal reciprocity and results in such very asymmetric transmission is the composite optical component of a linear polarizer combined with a quarter-wave plate that is at 45$^\circ$ to the axis of the polarizer.   Independent of the initial polarization such a device produces linearly polarized light in transmission in one direction and circularly polarized light in the other.   The Jones matrix for such a composite device with the linear polarizer on the input side would be 

\begin{equation}
\frac{1}{\sqrt{2}}\left[\begin{array}{cc}1 & i \\ 1 & i \end{array}\right] .
\end{equation}

Using Eqs. \ref{Tmatrix} and \ref{eigenvalues} one can see that this matrix is non-normal with non-orthogonal eigenpolarizations that are linear (45$^\circ$) and circular (L) polarized light associated with eigenvalues $1+i$ and 0 respectively.   Note that upon consideration of the backwards going $\hat{T} $ matrix one has the same pair of eigenvalues $1+i$ and $0$, but that their association has in a sense swapped e.g. they are associated with circular (R) and  linearly (45$^\circ$) polarized light respectively.  Such a device is still reciprocal in the sense of de Hoop as the forward going transmission matrix is equal to the transpose of the backwards going one.   This shows the central importance of the de Hoop condition as such a device obviously does not break time-reversal symmetry defined in any useful sense.

Based on our considerations of Sec. \ref{DiscreteSym} the highest symmetry materials that break diagonal reciprocity can have is $C_2$ around $z$.   Generally one finds that axes of dominant dichroism and birifrigence have to be non-aligned so that this kind of asymmetry may be most apparent in very anisotropic materials at frequencies of strong absorption.  Wave propagation of polarized light, but not the explicit symmetry properties, in crystals with nonorthogonal eigenpolarizations has been investigated previously \cite{Pancharatnam55a,Pancharatnam57a,Kushnir95a,Berry03a}.   Non-reciprocity of this kind has been also called eigenwave non-reciprocity in the literature and identified with a TRS breaking of a certain fashion \cite{Malinowski96a}.   However, it is important to emphasize that although materials that are non-reciprocal in this sense may have very asymmetric interaction with light in time and space, they don't themselves violate TRS.   For instance effects of this kind are purported to exist in zinc-blende type semiconductors, which break inversion, but obviously not TRS \cite{Zheludev94a}.   Because they are not constrained by any useful symmetry, materials that show a violation of diagonal reciprocity will typically have eigenstates that depend strongly on frequency.

Yet another (very) restricted form of reciprocity encountered in a typical experimental geometry when one ``flips" a sample to perform a transmission experiment from the backside.   If again de Hoop reciprocity is obeyed then the transmission matrix can be arrived at by using Eq. \ref{deHoopCondition} for backside transmission and performing a  $x \rightarrow -x$ transformation from the $\pi$ rotation around around the $y$ axis $a$ $la$ Eq. \ref{recip}.   

\begin{equation}
\hat{R}_x^{-1}(\pi) \hat{T}  \hat{R}_x(\pi) = \left[\begin{array}{cc}A & -C \\-B & D \end{array}\right].
\end{equation}

If this Jones matrix is identical to the forward going one then one will have the following constrained form for the forward going Jones matrix.

\begin{equation}
\left[\begin{array}{cc}A & B \\ -B & D \end{array}\right] .
\end{equation}

A comparison with Eq. \ref{C2x} shows that chiral states that obey de Hoop reciprocity are also reciprocal in this sense.   Materials which have only a mirror symmetry (Eq. \ref{reflection1}) will only be considered reciprocal in this sense if they are ``flipped" along a mirror axis.  I call this reciprocity $naive$ $reciprocity$.

Although these more restrictive formulations such as unitary, Hermitian, and diagonal reciprocity may have utility in analyzing experimental results, the most interesting cases of non-reciprocity will occur in situations when the most generic condition of de Hoop reciprocity is violated.   This will be the case for instance in situations where time-reversal symmetry is broken in the form of ferromagnetism, $p$-wave superconductivity (likely present in Sr$_2$RuO$_4$ \cite{Mackenzie03a}), magnetoelectrics \cite{Hornreich68a,Krichevtsov93a,Krichevtsov96a,Orenstein11a}, chiral spin liquids \cite{Machida09a,Ko11a} or in some proposed recent  \cite{Pershoguba13a,Orenstein11a,Aji12a}or older \cite{MarchRussell88a,Wen89a,Halperin89a,Levi91a, Dzyaloshinskii91a,Lawrence92a,Canright92a,Canright92b,Halperin92a,Shelankov92a} models for the pseudogap state of the high-temperature superconductors.

\section{Breaking of symmetries}
\label{Breaking}

As established in Secs. \ref{DiscreteSym} and \ref{TimeReversal}, particular symmetries  establish algebraic properties of or constraints on the Jones matrices.  It it is then straightforward to determine what combination of matrix elements are a measure of what particular symmetry breaking.   The methodology is clear.  Given a symmetry and the set of constrained Jones matrices for forward and reversed propagation that results from it, by inspection one establishes what sum or difference of matrix elements will yield identically zero if a particular symmetry is present.  If one establishes that the particular quantity is non-zero, this is a sufficient condition to establish that that symmetry is broken.

As a first example of this consider a tetragonal initially non-magnetic material that goes undergoes a symmetry breaking transition.   In the high symmetry state,  the material has all mirror symmetries and symmetry under  $\hat{R}_z(\pi/ 2)$  rotation.  The Jones matrix is then the simplest possible e.g.

\begin{equation}
\hat{T} = 
\left[\begin{array}{cc}A & 0 \\ 0 & A \end{array}\right] .
\label{tetragonal}
\end{equation}

If the phase transition is to a non-magnetic orthorhombic phase such that $\hat{R}_z(\pi/ 2)$ symmetry is broken, but the $x$ and $y$ mirror planes are maintained the Jones matrix will have the form of Eq. \ref{reflection} with diagonal components unequal and off-diagonal components still zero.  Therefore a measure of the orthorhombic order parameter is the quantity $|T_{xx} - T_{yy}|$.  This relation rests on the assumption that the laboratory $x-y$ frame coincides with the orthorhombic axes.   It is useful to have a more general relation.  Inspection of Eq. \ref{reflection1} and the relations that follow for the matrix elements in an arbitrary primed reference frame $x'-y'$, show that the quantity $|T_{x'x'} - T_{y'y'} + 2 i T_{x'y'}|$  will be sensitive to the symmetry breaking $|T_{xx} - T_{yy}|$ in the orthorhombic reference frame.

If the phase transition is to a  non-magnetic chiral phase such that all mirror symmetries are broken, but $\hat{R}_z(\pi/ 2)$ symmetry is preserved then the Jones matrix will have the form of Eq. \ref{C4} with equal diagonal components and off-diagonal components of identical magnitude and opposite sign.   Therefore a measure of the chiral order parameter is the quantity $|T_{xy} -T_{yx}|$.   This quantity will be finite if a different transmission coefficient is experienced for $R$ and $L$ circularly polarized light, but zero otherwise.   For instance it will be zero  for a linearly birefringent material that gives an effective rotation of a plane of polarized light if the polarized light is not incident along a mirror symmetry.

 If the transition is to a state that breaks macroscopic time-reversal symmetry (for instance to a ferromagnetic or magnetoelectric phase) the measure of macroscopic symmetry breaking and the order parameter becomes $|T_{xy} - T_{xy}^{b,+z}|$ where again the $T_{yx}^{b,+z}$ refers to backside transmission but in a sample that has been flipped.  Although it will also have a Faraday rotation, this quantity will be zero for a TRS chiral system.

\section{Combined symmetries and examples}
\label{Combined}

The various point group symmetries (reflection, rotation, rotation-reflection, inversion) and time reversal symmetry can be combined to get more insight into various broken symmetry states.    The simplest example of a TRS breaking state is a ferromagnet with its magnetization vector pointing along the propagation direction.   Both the time reversal operation and a $\hat{R}_x(\pi)$ rotation give the same state with the magnetization reversed.   Using Eq. \ref{TRScondition}, the time reversed Jones matrix is 

\begin{equation}
\bar{\hat{T}}^{+z}  = 
\left[\begin{array}{cc}T_{xx}^{b,-z} & T_{yx}^{b,-z} \\ T_{xy}^{b,-z} & T_{yy}^{b,-z} \end{array}\right] .
\end{equation}

The Jones matrix under $\hat{R}_x(\pi)$ rotation is

\begin{equation}
\hat{T}^{b,+z} = \hat{R}_x^{-1}(\pi)\hat{T}^{b,-z} \hat{R}_x(\pi) = 
\left[\begin{array}{cc}T_{xx}^{b,-z} & -T_{xy}^{b,-z} \\ -T_{yx}^{b,-z} & T_{yy}^{b,-z} \end{array}\right] .
\label{Rotated}
\end{equation}

In order that these expressions are equivalent, the Jones matrix must have the form

\begin{equation}
\hat{T} = 
\left[\begin{array}{cc} A &B \\ -B & D \end{array}\right] .
\end{equation}

If the material is cubic or tetragonal with $A=D$, the matrix is diagonalizable in the circular representation, otherwise the polarization eigenstates are elliptical.   In either case, this is consistent with a rotation of transmitted linearly polarized light e.g. a Faraday rotation.

A very interesting application of these methods is to the case of magnetoelectrics.   Consider the case of Cr$_2$O$_3$ which is a material that forms in the corundum structure with space group $D^6_{3d}$.   Below T$_N = 306$ K it becomes an commensurate antiferromagnet with the spins of the 4 Cr$^{+3}$ atoms per unit cell pointing along a three fold axis in an up-down-up-down alternating manner with magnetic point group $\bar{3}'m'$.   This structure breaks both inversion (P) and TRS (T), but the product of the two is a symmetry.   There are two degenerate ground states that are related to each other by either inversion or time-reversal.   As pointed out originally by Dzyaloshinskii \cite{Dzyaloshinsky58a} the magnetic symmetry of Cr$_2$O$_3$  allows linear magnetoelectric effects \cite{Astrov60a,Brown63a,Hornreich68a,Pisarev91,Krichevtsov93a,Krichevtsov96a,Fiebig05a} e.g. an electric field can cause a magnetization (with proportionality $\alpha_{ij}$) and a magnetic field can cause a polarization.  Such nonzero magnetoelectric coefficients are a consequence of so-called  ``$PT$" symmetry.  They are indicative of  inversion ($P$) and TRS ($T$) symmetries broken separately, but preserved under the combined $PT$ symmetry.   $PT$ symmetric situations where the $P$ symmetry breaking is such as to support a chiral state, have been called ``false chiral" by \textcite{Barron04a}.   

Magnetoelectrics can admit a number of  very interesting phenomenon including non-reciprocal gyrotropic birefringence, whereby there is a symmetric contribution to the dielectric tensor that rotates the axes of birefringence away from the crystal symmetry directions \cite{Brown63a,Hornreich68a}.   The sign of this contribution depends on the propagation vector and therefore the speeds of propagation can be different for counter-propagating beams.   Some evidence for this effect has been seen in transmission in continuous wave THz-range spectroscopy in Cr$_2$O$_3$ \cite{Mukhin97a} and related effects in reflection at higher optical frequencies \cite{Krichevtsov93a,Krichevtsov96a}.  The case of transmission is easier to analyze.   Consider the transmission along a direction perpendicular to the 3-fold axis and with the $x-y$ reference frame of the linearly polarized light aligned to the structural axes. The  $\hat{T}$ matrix transformed under inversion $\hat{P}$ is

\begin{equation}
\hat{T}^{b,+z} = \hat{P}^{-1}\hat{T}^{b,-z} \hat{P} = 
\left[\begin{array}{cc}T_{xx}^{b,-z} & T_{xy}^{b,-z} \\ T_{yx}^{b,-z} & T_{yy}^{b,-z} \end{array}\right] .
\end{equation}

The same state should be obtained under time-reversal

\begin{equation}
 \overline{ \hat{T} }  =  \left[\begin{array}{cc}T_{xx}^{b,-z} & T_{yx}^{b,-z} \\ T_{xy}^{b,-z} & T_{yy}^{b,-z} \end{array}\right] .
\end{equation}

The  $\hat{T}$ matrix must therefore have the form 

\begin{equation}
\hat{T} = 
\left[\begin{array}{cc} A &B \\ B & D \end{array}\right] .
\label{gyro}
\end{equation}

No other symmetries further constrain this matrix and so only equal off-diagonal entries are admitted, and hence it can be diagonalized by a rotation.   As I have already assumed in my analysis that the coordinate axes themselves are aligned along the principle structural axes of the system this shows that the optical axes of the linear dichroism/birefringence can be rotated from the coordinate system.  This is consistent with the phenomena of gyrotropic birefringence.  We can get further insight into this system by realizing that the material is invariant under a $\hat{R}_y(\pi)$ rotation.   The form of the Jones matrix is then of Eq. \ref{Rotated}, which determines that the off-diagonal components for forward and backwards propagation are opposite to each other showing the sense of rotation of the optical axes is opposite for front and back surfaces, giving evidence for the non-reciprocal nature of this effect.   Moreover, since the  $\hat{T}$ matrix has the form of Eq. \ref{gyro}   $\hat{R}_x(\pi)$ symmetry one can probe the non-reciprocal nature of the state by measuring  $T_{xy} $ on a single surface alone.   Note that presence of TRS in the paramagnetic state constrains (through the reciprocal condition Eq. \ref{deHoopCondition}) the off-diagonal components of the $\hat{T}$ matrix to be zero above T$_N$.

For transmission along the $c$-axis the form of Eq. \ref{gyro} is still valid because it was based on the combined symmetry $PT$  without regards to propagation direction.  However Cr$_2$O$_3$ has a three fold axis along $z$ and hence its $\hat{T}$ matrix must be compatible with the constraints of the $C_3$ symmetry given in Eq. \ref{C3} for propagation along $z$.   This gives a $\hat{T}$ matrix with zero off-diagonal components and identical diagonal ones.  Although its been shown that magnetoelectrics like Cr$_2$O$_3$ show a non-reciprocal Kerr rotation effect in reflection \cite{Hornreich68a,Krichevtsov93a,Krichevtsov96a}, symmetry constraints determine that the transmission of a single crystal slab is invisible to such non-reciprocal effects e.g. no Faraday effect is exhibited \cite{Dzyaloshinskii95a}. Microscopically one can consider that the light suffers a rotation of order $\alpha$ (where $\alpha$ is the mangetoelectric coefficient) upon crossing the front surface in transmission, propagates through the sample with $E$ and $H$ fields at a small angle deviating from 90$^\circ$ by $\alpha$, and then undergoes a rotation with the same magnitude at the back surface, but in the opposite direction.   Note that inversion symmetry breaking by a substrate will invalidate this cancelation and it should be possible to see a Faraday rotation $\phi_F \propto (n_s - 1) \alpha$ in transmission for Cr$_2$O$_3$ on a substrate of index of refraction $n_s$ \cite{Shelankov92b}.  An analysis of the reflection coefficient (discussed below) for a TRS breaking state such as this one show that it can exhibit a Kerr rotation.

Even more exotic states in the form of various chiral spin liquid states can be analyzed \cite{Simonet12a}.   A ``vector spin chiral" state is said to be exhibited when the quantity  $   \langle  S_i \times S_j \rangle   \neq 0 $, while  $  \langle  S_i  \rangle  =0$ \cite{Katsura05a,Kenzelmann05a}.  Such a state does not break TRS and despite its accepted name, is not necessarily chiral in a rigorous sense.  By itself the finite expectation value is not enough to determine whether such a state is chiral and hence specify its electrodynamics.   The related quantity $   \langle e_{ij} \cdot S_i \times S_j \rangle  \neq 0   $, where $e_{ij}$ is a unit vector connecting sites $i$ and $j$, does break inversion and all mirror symmetries and so such an order parameter has a definite handedness.  By the considerations of section \ref{DiscreteSym} this state should have the transmission properties of other chiral systems and can exhibit a Faraday rotation (but not a Kerr rotation, see Sec. \ref{extension}).

Other vector spin chiral states may or may not exhibit polarization anisotropies depending on the details of their ordering.   For instance, a magnetic cycloid chain, consisting of coplanar spins that rotate around a perpendicular axis as one moves along an axis parallel to the spin plane has  $   \langle  S_i \times S_j \rangle   \neq 0 $, but is achiral.  Two different senses of rotations of the spins can be generated by the inversion operation, but these states are not uniquely specified by $   \langle  S_i \times S_j \rangle   $,  as they can be interconverted by a $\hat{R}_z(\pi)$ rotation.   Since we have already shown that a $\hat{R}_z(\pi)$ rotations have no effect on transmission, a cycloid order of this form cannot be detected optically.

A ``scalar spin chiral" state is said to be exhibited when the quantity  $   \langle S_i \cdot S_j \times S_k \rangle   \neq 0 $, while  $  \langle  S_i  \rangle  =0$ \cite{Wen89a,Machida09a,Ko11a}.     A state which exhibits this order with the same sign on a line along the propagation direction  (e.g. ``ferro" ordering) breaks time-reversal symmetry, but not inversion.  It will show both Kerr and Faraday reflection.   As pointed out in Ref. \onlinecite{Dzyaloshinskii91a} states which exhibit an ``antiferro" ordering (between say CuO$_2$ planes in the cuprates as considered historically \cite{Wen89a,Halperin92a}) and break inversion structurally will show a Kerr rotation (but not Faraday) in a manner very similar to the case of Cr$_2$O$_3$.

\section{Extension of Jones matrix formalism to other geometries and experiments}
\label{extension}

With some modifications the above arguments can be extended to other geometries and experiments.   In reflection the loss of symmetry across the free surface means particular consideration must be given.  Although the point group symmetry constraints applied to the transmission matrix when only a single surface is considered apply equally to the normal incidence reflection matrix, there can be additional issues that arise due to the lack of symmetry at the interface\cite{Vernon80a}.  Moreover the de Hoop reciprocity relation for TRS I found above was derived specifically for the case of transmission. Similar constraints on reflection require different consideration.

A practical matter that arises when considering reflection, is the long standing controversy regarding whether or not a Kerr-like rotation can occur in reflection from a chiral TRS media \cite{Casimir66a,Bungay93a,Halperin92a,Bahar09a,Hosur14a,Cho14a}.   A straightforward application of Fresnel's equation using the bulk (antisymmetric) dielectric constant predicts that a Kerr rotation should be observed if there is finite dissipation \cite{Landau60a}.   However it is clear that one cannot simply export the bulk dielectric constants to the surface as their antisymmetry arises from an expansion of the dielectric constant in odd powers in the photon wavevector $k$.   Such an expansion cannot be strictly valid at a surface and one must retain the full spatial dependence of the dielectric constant.   Moreover,  in chiral systems, the issue of boundary conditions is complicated by a certain freedom in choice of $\epsilon$ and $\mu$.  Gyrotropy (e.g. dependence of dielectric constant on $k$) can be included only in $\epsilon$ or in $\mu$ or both.   There is a disagreement within different calculational schemes, as some observable quantities, such as reflectivity, turn out to be dependent on these choices.   For instance an alternative treatment using essentially magnetoelectric-like ``Casimir" material relations \cite{Casimir66a}, finds that no rotation should be occur.   Other work says that both approaches are essentially valid and correctly applying terms involving derivatives of the non-local susceptibilities gives rotation in reflection \cite{Bungay93a}.  Even more recent work concludes that the differential reflection upon normal incidence should vanish \cite{Bahar09a}.

The present formalism however provides a simple proof that materials that preserve TRS cannot exhibit a Kerr rotation.  First, one realizes that in the simplest normal incidence reflection geometry with source and detector spatially coincident, the ``forward" going reflection matrix must be the same as the ``backward" going reflection matrix.  Switching the source and the detector gives no change to any physical quantity, because ``forward" going and ``backward" going are the same thing.   Therefore a similar treatment of de Hoop reciprocity leading to Eq. \ref{TRScondition} applied to the reflection matrix shows that the time-reversed reflection matrix will be the transpose of the reflection matrix for the forward direction in time e.g.

\begin{equation}
\overline{ \hat{R}}   =    \hat{R}^T.
\label{TRSconditionReflection}
\end{equation}

\noindent It follows that if the material has TRS and hence the reflection matrix doesn't change under the time-reversal operation (e.g. $     \hat{R}  =  \overline{ \hat{R}} $), then the off-diagonal components of the reflection matrix must be equal.   Therefore a material with TRS cannot exhibit a Kerr rotation as that necessitates anti-symmetric off-diagonal components in the reflection matrix.   A somewhat related, but much more involved proof, for generic geometry can be found in Ref. \onlinecite{Halperin92a}.   The absence of a Kerr effect except in the presence of TRS breaking is relevant for the interpretation of recent experiments on cuprate superconductors that have seen a Kerr rotation onset in underdoped compounds \cite{Xia08a,Karapetyan12a,Karapetyan13a}.   Such experiments have been interpreted as being consistent with the onset of a chiral phase, but this appears to not be a possible explanation (at least at the level of linear response and in thermal equilibrium).   See also \textcite{Hosur012a,Hosur14a,Cho14a} for further discussion.

Most optical experiments do not measure complex transmission coefficients directly.   For experiments like Fourier transform infrared reflectivity (FTIR) \cite{Basov05a,Basov11a},  Sagnac interferometry \cite{Kapitulnik09a}, or spectroscopic ellipsometry \cite{Azzam78a} the light intensity is detected.   In this case one has to make use of the Mueller-Stokes matrix formalism, where the optical resonse is given by a real $4 \times 4$ ``Mueller" matrix operating on a real 4 component ``Stokes" vector.   The Stokes vector is given in terms of the electric field intensities in different directions and polarizations e.g.  

\begin{equation}
\left[\begin{array}{c} |E_{x}|^2 +   |E_{y}|^2  \\   |E_{x}|^2 -   |E_{y}|^2  \\  |E_{45^\circ}|^2 -   |E_{-45^\circ}|^2  \\  |E_{l}|^2 -   |E_{r}|^2 \end{array}\right].
\label{Stokes}
\end{equation}

\noindent The Mueller matrix represents the intra- and interconversion of these polarization states.   The Mueller matrix formalism is more general than the Jones formalism as its reliance on measured intensities means that it can characterize non-coherent unpolarized light.   Most relevant experiments however can utilize polarized light and in this case  it is possible to convert from the Jones to Mueller matrices for analysis.   This then suggests a method for analyzing the symmetry properties of materials being probed by such a technique;   analyze their symmetry properties and apply the relevant constraints to the simpler $2 \times 2$ Jones matrices first and then convert the Jones matrices to Mueller matrices.   This method should be completely valid as long as the sample under test is not depolarizing (which is typically the case).

The Jones-to-Mueller matrix conversion is performed by calculating the light intensities from the electric fields.   Following Azzam and Bashara \cite{Azzam78a} and starting from the Jones formalism, which gives  $\hat{T}  E_i = E_t$ I begin by calculating 
 
\begin{equation}
E_t  \otimes E_t^*= \hat{T}  E_i   \otimes  \hat{T} ^* E_i^*  =    ( \hat{T}   \otimes \hat{T} ^*)(     E_i   \otimes   E_i^*).
\end{equation}

\noindent where the $E_t  \otimes E_t^*$ and  $\hat{T}  \otimes \hat{T} ^* $ are the direct products yielding a 4 row vector and a  $4 \times 4$ matrix respectively.   The $E$ field products yields the  ``coherence" vector.

\begin{equation}
C = E \otimes E =  \left[\begin{array}{c} E_{x}  E_{x}^*   \\  E_{x} E_{y}^*   \\  E_{y} E_{x}^*  \\ E_{y} E_{y}^*  \end{array}\right].
\label{coherencyvector}
\end{equation}

Therefore 

\begin{equation}
C_t =  (\hat{T}  \otimes   \hat{T} ^*)  C_i.
\label{coherencyvector2}
\end{equation}

By inspection of Eq. \ref{Stokes} and Eq. \ref{coherencyvector} one can express the Stokes vector in terms the coherency vector as

\begin{equation}
S = \hat{A} C
\label{coherencyvector3}
\end{equation}

where 

\begin{equation}
 \hat{A} =  \left[\begin{array}{cccc}1 & 0 & 0& 1  \\  1 & 0 & 0& -1  \\   0 & 1 & 1& 0  \\  0 & i & -i & 0   \end{array}\right] .
\label{coherencyvector4}
\end{equation}

Substituting into  Eq. \ref{coherencyvector2} the inverse of Eq. \ref{coherencyvector3}, we get

\begin{equation}
S_t = \hat{A} (\hat{T}  \otimes  \hat{T}^*)  \hat{A} ^{-1}  S_i .
\end{equation}

So therefore in the Mueller matrix expression   $S_t =  \hat{M}  S_i$ the Mueller  matrix $ \hat{M} $ is given by $    \hat{M}  =  \hat{A}  (\hat{T}  \otimes  \hat{T}^* )  \hat{A} ^{-1} $.   Using the direct product expression for $ \hat{T} \otimes  \hat{T} ^*$ and the transformation $\hat{A} $ and its inverse one can express the Mueller matrix $ \hat{M} $ in terms of the components of the Jones matrix (as defined in Eq. \ref{ABCD}) as 

\begin{widetext}
\begin{equation}
\normalsize
\left[\begin{array}{cccc} \frac{1}{2} (AA^* +   BB^*  +   CC^* + DD^*    )  & \frac{1}{2} (AA^* -   BB^*  +   CC^* - DD^*    )  &  Re(A B^* )  +  Re(C D^* )      & - Im(A B^* )  -  Im(C D^* )  \\  \frac{1}{2} (AA^* +  BB^*  -   CC^* - DD^*    ) & \frac{1}{2} (AA^* -   BB^*  -   CC^* + DD^*    ) &Re(A B^* )  -  Re(C D^* )  & - Im(A B^* )  +  Im(C D^* )   \\   Re(A C^* )  +  Re(B D^* )    &  Re(A C^* )  -  Re(B D^* )   &  Re(A D^* )  +  Re(B C^* )   &  - Im(A D^* )  -  Im(B C^* )  \\   Im(A C^* )  + Im(C D^* ) &Im(A C^* )  - Im(C D^* ) & Im(A D^* )  + Im(B C^* ) & Re(A D^* )  -  Re(B C^* )   \end{array}\right] .
\vspace{5mm}\label{MuellerMatrix}
\normalsize
\end{equation}
\end{widetext}

The symmetry constraints on the Jones matrices discussed in Secs. \ref{DiscreteSym} and \ref{TimeReversal} can be directly applied to Eq. \ref{MuellerMatrix} and with a great deal of algebra give constraints on the Mueller matrices.

\section{Conclusions}

In this work I discuss a Jones transfer matrix formalism for inferring the existence of exotic broken symmetry states of matter from their electrodynamic response.  This formalism is directly applicable to time-domain THz spectroscopy in transmission, but is more broadly applicable.  I discussed the consequences of discrete broken symmetries on ordered states of matter including the presence and absence of reflections, rotations, inversion, rotation-reflections and  time-reversal symmetries and the constraints they give on Jones matrices.   These constraints typically appear in the form of an algebra relating matrix elements or overall constraints (transposition, unitarity, hermiticity, normality, etc.) on the form of matrix.  As usual, the utility of symmetries is that one can still deduce quantitative and qualitative information even when the underlying equations of motion are unknown.   A number of explicit examples were given that are of current relevance to the study of correlated broken symmetry states of matter.   One important consequence of this formalism is the demonstration that Kerr rotation must be absent in time-reversal symmetric chiral materials.  

\section{Acknowledgments}

I would like to thank R. Valdes Aguilar,  N. Drichko,  A. Fried, A. Kapitulnik, S. Kivelson, Y. Lubashevsky, C. Morris, L. Pan, S. Raghu, O. Tchernyshyov, A. Turner, C. Varma, and V. Yakovenko for useful discussions on related topics.   This research was supported by the Gordon and Betty Moore Foundation through Grant GBMF2628 and the DOE-BES through DE-FG02-08ER46544.

\bigskip

\bigskip

\bibliography{HighTcRef2}

\end{document}